\documentclass[a4paper,12pt]{article}
\usepackage{amssymb}
\usepackage{colortbl}
\usepackage{amstext}
\newcommand{\be}{\begin{equation}}
\newcommand{\ee}{\end{equation}}
\newcommand{\bea}{\begin{eqnarray}}
\newcommand{\eea}{\end{eqnarray}}
\def\bra{\langle}
\def\ket{\rangle}

\title{%
  {\bfseries Proposal for a CFT interpretation of}\\ 
  {\bfseries Watts' differential equation for percolation}
}
\author{%
  {\normalsize\sc Michael Flohr\thanks{{\tt flohr@th.physik.uni-bonn.de}, 
    research supported by the European Union Network HPRN-CT-2002-00325 
    (EUCLID)}}\ \ \ \ {\normalsize and}\ \ \
  {\normalsize\sc Annekathrin M\"uller-Lohmann\thanks{{\tt 
    anne@th.physik.uni-bonn.de}}}\\[0.5cm]
  {\normalsize\slshape Physikalisches Institut}\\[-0.1cm]
  {\normalsize\slshape University of Bonn}\\[-0.1cm]
  {\normalsize\slshape Nussallee 12}\\[-0.1cm]
  {\normalsize\slshape D-53115 Bonn, Germany}
}
\date{{\small\today}}
\begin{document}

\maketitle

\begin{abstract}
{\small G.$\,$M.$\,$T.~Watts derived \cite{Watts:1996yh} that in 
two dimensional critical percolation the crossing probability $\Pi_{h v}$ 
satisfies a fifth order differential equation which includes another one of 
third order whose independent solutions describe the physically relevant 
quantities $1,\Pi_h, \Pi_{hv}$. 

We will show that this differential equation can be derived from a level 
three null vector condition of a rational $c=-24$ CFT and motivate how
this solution may be fitted into known properties of percolation.}
\end{abstract}

\begin{picture}(0,0)
  \put(300,440){{\tt \ hep-th/0507211}}
  \put(300,451){{\tt BONN-TH-2005-02}}
\end{picture}

\section{A brief review of percolation properties}

According to Langlands et al \cite{Langlands:1994}, critical percolation 
in two dimensions has interesting features in conformal field theory such 
as the conformal invariance of the three independant crossing probabilites 
$1,\Pi_h, \Pi_{hv}$. For $\Pi_h$, Cardy \cite{Cardy:1991cm} was already 
able to derive an exact solution with the help of boundary conformal field 
theory which matches the numerical data to a high accuracy. Starting from 
this background, Watts \cite{Watts:1996yh} motivated how $\Pi_{hv}$ can be 
expressed by a correlation function of boundary operators in the 
$Q \rightarrow 1$ limit of the $Q$-state Potts model and deduced a differential 
equation of fith order that agrees with the simulations. Additionally he 
observed that the three physically relevant solutions already satisfy a 
third order differential equation. We will not give a review on percolation
here, for details see, e.g., Kesten \cite{Kesten} or Stauffer and Aharony
\cite{Stauffer} instead.

In the previous literature, several arguments have been given to describe 
the crossing probabilities in two dimensional critical percolation as 
conformal blocks of a four point correlation function of $(h=0)$-operators 
in a $c=0$ conformal field theory (CFT), using a second (third) level null
 vector to get $\Pi_h$ ($\Pi_{hv}$). The most prominent are 
\begin{itemize}
\item[(1.)] (for $c=0$) the Beraha numbers 
$Q = 4 \cos^2 \left(\frac{\pi}{n}\right)$ (with $n$ usually denoted as 
$m+1=2,3,4 \ldots$ which in most Potts models are related to the central 
charge by $c = 1-\frac{6}{m(m+1)}$ \cite{Cardy:1991cm});
\item[(2.)] (for $c=0$) the differential equation for $\Pi_h$ can as well 
be derived by the Stochastic/Schramm Loewner Evolution (SLE) which 
strengthens the first argument;
\item[(3.)] (for $h=0$) the ratio of the partition functions for free 
boundary conditions to $Z=1$ of percolation (as suggested by Cardy 
\cite{Cardy:1991cm}); 
\item[(4.)] (for $c=h=0$) the interpretation of the central charge as 
describing the finite size effects of the energy.
\end{itemize}

To understand the first point, we give a brief review on the $Q$-state 
Potts model (literature for the connection to percolation can be found in
\cite{Hu1,Hu2,Hu3,Hu4,Potts}). On a simply connected compact region with a piecewise 
differentiable boundary the horizontal crossing probability $\Pi_h$ is 
defined through the partition function. It has originally been derived
by Fortuin and Kasteleyn \cite{Fortuin,Fortuin2} but can also be looked up in,
e.g.,
the literature given above or \cite{Cardy:1991cm,Kleban:1999rw,Wu:1982ra}. 
\begin{equation}
Z = \prod_{(r,r')} \left( 1 + x \delta_{s(r),s(r')}\right) = \sum_G Q^{N_c} x^{N_b},
\end{equation}
where $x=\frac{p}{1-p}$ for $Q \rightarrow 1$ and the rightmost sum running 
over all possible graphs of $N_b$ bonds in $N_c$ clusters. By expanding it 
in powers of $x$ we can extend the $Q$-state Potts model to $Q \in \mathbb{R}$.

$\Pi_h$ describes the probability of having a connection from, e.g., one piece 
$X=(x_0,x_1)$ of the boundary to another disjoint part $Y=(x_2,x_3)$ where 
the spins are fixed to values $\alpha$ and $\beta$, respectively, while on 
the rest we have free boundary conditions (for a more detailed introduction 
see \cite{Cardy:2001vr}). Hereby any region which can be mapped onto the real 
axis by a conformal transformation is equivalent (for corners we may get singular 
behavior but no discontinuities at the corresponding points). For 
$\alpha \neq \beta$, it is given by \cite{Kleban:1999rw}
\begin{equation}
\Pi (X,Y) = \lim\limits_{Q \rightarrow 1} \left(1 - \frac{Z_{\alpha \beta}}{Z_{\alpha \alpha}} \right).
\end{equation}
In terms of boundary changing operators \cite{Cardy:1989,Belavin:1984vu} 
from free ($f$) to fixed ($\alpha,\beta$) 
conditions, we get
\begin{equation}
Z_{\alpha,\beta} = Z_f \langle \phi_{(f|\alpha)} (x_0)\phi_{(\alpha|f)} (x_1)\phi_{(f|\beta)} (x_2)\phi_{(\beta|f)} (x_3)\rangle .
\end{equation}
In the infinite volume limit, these quantities diverge  for $Q \neq 1$, but by 
taking a closer look at the partition function of the Potts Model for 
$Q \rightarrow 1$, we find for a minimal model with central charge $c=0$ the 
partition function to be $Z=1$ in this limit. 

For $\Pi_h$, the $\phi$ are $h_{(1,2)}$ boundary operators, while the results for 
$\Pi_{hv}$ contain other boundary operators that can be identified by comparison 
with known Potts models (i.e. for $Q=2,3$) to have weight $h_{(1,3)}$. Another 
motivation for this ansatz can be found by letting the length of the segment 
with free boundary conditions tend to zero. Therefore we know from fusion rules, 
that
\begin{equation}
 \phi_{(\alpha|f)} \times \phi_{(f|\beta)}  \sim \delta_{\alpha \beta} + \phi_{(\alpha|\beta)}
\end{equation}
which means that the fusion of two $\phi_{(1,2)}$ boundary operators yields a
$\phi_{(1,3)}$ field (see Cardy \cite{Cardy:1991cm}, Kleban \cite{Kleban:1999rw}).
Hence we will look out for a rational CFT with a Kac table which is large
enough to contain level three fields (i.e. $\phi_{(1,3)}$ or $\phi_{(3,1)}$).

So far, it seems very reasonable to choose $c=0$ to describe percolation, but, 
unfortunately, a minimal model $c_{(3,2)} = 0$ is not very interesting, since 
its field content only consists of two $h=0$ fields -- $\phi_{(1,1)}$ and 
$\phi_{(1,2)}$. Thus the $Q \rightarrow 1$ limit of the $Q$-state Potts Model 
(which corresponds to $c_{(3,2)} = 0$ since both partition functions equal one)
does not accomodate Cardy's proposal that boundary operators for the 
horizontal vertical crossing probability should appear at level $rs = 3$ in the 
Kac table. Thus we might wish to not follow his original approach to the 
horizontal crossing probability but to reconsider our underlying CFT.

In fact, if we include the $\phi_{(1,3)}$ field into the spectrum of our
conformal field theory with vanishing central charge, the partition function
will not be equal to one. More precisely, including this field with conformal
weight $h_{(1,3)}=1/3$ into the spectrum leads to a logarithmic conformal
field theory, see \cite{Flohr:2001zs,Gaberdiel:2001tr,Gurarie:1993xq} and
references therein. 
The representation with this conformal weight is indecomposable,
containing an irreducible sub-representation with character
\begin{equation}
  \chi_{(1,3)}^{}(q) = 
  \frac{1}{\eta(q)}\sum_{n\in\mathbb{Z}}(2n+1)q^{3(4n+1)^2/8}\,,
\end{equation}
where $\eta(q)$ denotes the Dedekind $\eta$-function $q^{1/24}\prod_{n\geq 1}
(1-q^n)$. This logarithmic conformal field theory is a so-called augmented
minimal model, and it is rational in the sense that it possesses only 
finitely many indecomposable or irreducible representations. However,
the resulting modular invariant partition function for this model is,
up to terms proportional to $\log(q\bar q)$, given by the partition
function of a $c=1$ theory\footnote{Note that the \emph{effective}
central charge of this model is $c_{{\rm eff}}=c-24h_{{\rm min}}=1$.} with
radius of compactification given by $2R^2=1/(2\cdot 3)=1/6$, namely
\begin{eqnarray}
  Z &=& \frac{1}{|\eta(q)|^2}\left(|\Theta_{0,6}(q)|^2 
     + 2\sum_{\lambda=1}^5|\Theta_{\lambda,6}(q)|^2
     + |\Theta_{6,6}(q)|^2\right)\,,\\
  \Theta_{\lambda,k}(q) &=& \sum_{n\in\mathbb{Z}}q^{(2kn+\lambda)^2/4k}\,.  
\end{eqnarray}
The logarithmic corrections cannot be fixed in magnitude by the requirement
of non-negative integer coefficients in their respective $q$-expansions,
but we mention for completeness that
\begin{eqnarray}\label{eq:Zfull}
  Z_{{\rm full}}[\alpha,\beta] &=& Z + \alpha\frac{\log(q\bar q)}{|\eta(q)|^2}
  \sum_{\lambda=1}^5
  |(\partial\Theta)_{\lambda,6}(q)|^2 + \beta\log(q\bar q)^2|E_2(q)|^2\,,\\
  (\partial\Theta)_{\lambda,k}(q) &=& 
  \sum_{n\in\mathbb{Z}}(2kn+\lambda)q^{(2kn+\lambda)^2/4k}\,,
\end{eqnarray}
and $E_2(q)$ is the Eisenstein series of modular weight two. Such
modular invariants can be found by solving the modular differential equation,
which must be satisfied by any finite-dimensional representation of the
modular group in terms of modular functions (with multiplicative systems).
Usually, it suffices to know one character of the conformal field theory,
e.g.~the vacuum character, and the spectrum, i.e.~the conformal weights
of all admissible irreducible or indecomposable representations.
Details on how this construction works in the case of logarithmic CFTs can be
found in \cite{Flohr:1995ea,Flohr:1996vc}.
In any case,
including the field $\phi_{(1,3)}$ from the boundary of the Kac-table of the
$c_{(3,2)}=0$ minimal model results in an enlarged theory with partition 
function definitely not being equal to one.

Now we will take a look at the second argument for $c=0$ from 
Stochastic/Schramm Loewner Evolution (SLE). SLE is based on the 
orignal work of Loewner \cite{Loewner} and has been applied to
Brownian motions, e.g.~by Lawler, Schramm, Werner and Rhode 
\cite{math/9911084,lawler-2001-187,math.PR/0106036,OdedSchramm}.
These random curves can provide us with another way to formulate
the percolation problem (various introductions can be found, e.g.,
in \cite{gruzberg-2004-114, LawlerCornell,encyclopedia,werner-2000-,werner-2003-}). 
Unfortunately, up to now
it has not been possible to establish a link between Dubedat's 
\cite{Dubedat:2004} proof for Watts' differential equation within
an SLE approach and a CFT bond percolation model. Thus we will 
concentrate on the results for the solution of Cardy's differential
equation in the following. Although the issues discussed above 
concerning the insufficient field content of the minimal model with $c=0$ 
do not apply within the SLE setting, we will show that SLE does not 
necessarily force us to take a CFT with vanishing central charge $c=0$. 

In \cite{Cardy:2005kh}, Cardy gave an elaborate review of how SLE can be 
applied to calculate crossing probabilities. Simply speaking, a path evolves by a Brownian 
motion of speed $\kappa =6$ which repeatedly hits the real axis. In a 
configuration where the motion starts from a point $a_0$ on the real axis 
running all over the complex upper half plane with $x_1 <a_0<x_2$ being the 
end points of the crossing intervals, one of the points will be ``swallowed'' 
first. For $x_1$ being the first to be hit by the graph, there obviously 
exists a free path along the outer line of the graph, for $x_2$ it is quite 
as obvious that this is not the case. Thus the probability that there is a 
crossing between $(a_0,x_2)$ to $(-\infty,x_1)$ is given by a Bessel process, 
described by a differential equation
\begin{equation}
\left( \frac{2}{x_1 - a_0} \frac{\partial}{\partial x_1} + \frac{2}{x_2 - a_0} \frac{\partial}{\partial x_2} + \frac{\kappa}{2} \frac{\partial^2}{\partial a_0^2}\right) P(x_1,x_2; a_0)\,.
\end{equation}
>From translational invariance we get $\partial a_0 = - \partial x_1 - \partial x_2$ 
and from conformal invariance, we know, that $P$ is a function of the ratio 
$\eta = \frac{x_2 - a_0}{x_1-a_0}$. This is exactly the same differential 
equation one obtains with CFT for percolation from a two level null vector 
\cite{Cardy:1991cm}. There is also a general expression, relating the speed 
of the Brownian motion $\kappa$ to the central charge and thus the highest 
weight states of the Virasoro algebra (i.e.~\cite{Bauer:2002tf}, \cite{Cardy:2005kh})
\begin{eqnarray}
c^\kappa &=& \frac{(3 \kappa - 8)(6 - \kappa)}{2 \kappa}\,,\\
h_{(r,s)}^\kappa &=& \frac{(r \kappa - 4 s)^2 - (\kappa - 4)^2}{16 \kappa}\,.
\end{eqnarray}
Hence, $c=0$ and $h_{(1,2)} = 0$ for $\kappa = 6$ which has been shown to describe 
$\Pi_h$ in two dimensional critical site percolation on the triangular lattice 
\cite{Smirnov:1994}. Additionally, as stated by
Bauer and Bernard \cite{Bauer:2002tf}, there is a direct correspondence 
between the $Q$-state Potts model and SLE
\begin{equation}
Q = 4 \cos^2 \left( \frac{4 \pi}{\kappa} \right)\,, \quad \quad \kappa \geq 4,
\end{equation}
by matching the known value of the dimension of the boundary changing operator 
for the $Q$-state Potts model with $h_{(1,2)}^\kappa$. 

The third argument makes use of the form of the partition function of 
the $c=0$ model. But as we already have shown, the partition function for 
the augmented $c=0$ model is not the same as for the minimal $c=0$ model 
and thus especially not equal to unity. From this argument, we will show, 
that we do not longer have to choose $h=0$ operators as suggested by Cardy 
\cite{Cardy:1991cm}. 

Regarding the problem mentioned above with only a single region with fixed 
boundary conditions, in the $Q \rightarrow 1$ limit, we have
\begin{equation}
Z_{\alpha} = Z_f \langle \phi_{(f|\alpha)} (x_0)\phi_{(\alpha|f)}(x_1)\rangle = Z_f \times (x_0-x_1)^{2h}.
\end{equation}
In the minimal model, both partition functions are equal to unity, thus $h=0$, 
but in the extended model, we do not know the exact form of $Z_f$, hence the 
boundary operator is not a priori fixed in its dimension.

The last point addresses the transformation back onto the original region that 
is described by the formula \cite{Cardy:1991cm}
\begin{equation}
\langle \phi_0(w_0) \phi_1(w_1) \ldots \rangle = \prod_i |w'(z_i)|^{-h_i} \langle \phi_0(z_0) \phi_1(z_1) \ldots \rangle.
\end{equation}
The expression has a physical meaning in the general non scale invariance of 
critical systems which picks up a factor $(L/L_0)6{a c}$ with $L$ being the 
overall size of the region, $L_0$ some non universal microscopic scale 
(i.e.~the lattice spacing), $c$ the (effective) central charge and $a$ being 
dependent on the geometry (i.e.~$a = -\pi / \gamma$ if the boundary operator 
sits in a corner with an interior angle $\gamma$, see 
\cite{Cardy:1991cm,Kleban:1992ws,Kleban:1999rw}). 
Since percolation is assumed to be scale invariant, 
the effect of the conformal mapping should vanish. But the physical properties 
of our system only depend on the differential equation arising from null vectors, 
thus this condition only has to hold in this sense. 

We remark here that the above argument of finite size scaling effects relies
on an analysis of the asymptotic behavior of the partition function. This
behavior, however, depends on the central charge only modulo $24$.
Moreover, invariance of the correlation functions holds in any conformal
field theory, as long as the Jacobian transformation factors are properly
accounted for. We will see below that within our proposal, where the
crossing probabilities are obtained from a CFT with non-vanishing central
charge, we have quotients of correlation functions such that the final 
expressions have all desired properties. 

Recapitulating, we state that the assumptions on percolation should be 
reconsidered, since most arguments do not seem to be as strict as stated 
before, i.e.~the central charge arguments most times refer to an effective 
central charge $c_{{\rm eff}} = c-24h_{{\rm min}}$ where $h_{{\rm min}}$ is 
the weight of the ground state. Thus $c_{{\rm eff}}>c$ in the case of 
non-unitary theories with negative weights. 
Thus, the arguments for $h=0$ are either problematic due to the $c=0$ minimal 
model being nearly empty or are connected with the central charge. 
Hence, once we agree on the proposal that we should work with the augmented, 
and therefore non-unitary, $c=0$ model, we also have to deal with the effective 
charge in that model -- which is the same for both the theories considered
in this work, 
\begin{equation}
  c_{(6,1)}=-24 \equiv 0=c_{(3,2)}\ \textrm{mod}\ 24\,. 
\end{equation}
\section{The Watts differential equation}
As already mentioned, Watts \cite{Watts:1996yh} derived a fifth order 
differential equation for $\Pi_{hv}$, starting from a $c=0$ theory with 
$h_{1,2}=0$ boundary changing operators following Cardy's ansatz for $\Pi_h$. 
A priori, as a minimal model $c_{(3,2)}=0$ we only have two primary
fields within the Kac table, the identity residing at $(1,1)$ and its
duplicated entry. Thus if we assume a null state on the first level
$L_{-1} | 0 \rangle$, we quickly see that from the generic form of 
the level two null state follows that $L_{-2} | 0 \rangle = 0$, too,
and so on, until the only non-vanishing state is the vacuum itself.
Thus, within a true minimal model, there can not be a `direct'' null 
vector on the fifth level whatsoever. Thus, when talking about
higher than level two null vectors in a $c=0$ rational CFT, we have
to add the note that by talking about $c=0$ we refer to the augmented minimal
model, i.e. $c_{(9,6)}=0$. Whether in this LCFT a null state on the
fifth level exists or not remains to be shown. Nevertheless, Watts
came up with the correct differential equation for the 
horizontal-vertical-crossing probability in percolation by motivating a level
five null vector which can be interpreted as a level three null
vector acting on a level two state as shown in \cite{Kleban:2002pf}.
In a $c=0$ theory, it seems strange, that in contrary to the results for 
$\Pi_h$, the $\Pi_{hv}$ boundary operators cannot be identified directly 
\cite{Kleban:1999rw}. Considering the asymptotic behavior, one can find the 
correct expressions for $\Pi_h$ and $\Pi_{hv}$ \cite{Kleban:2002pf} by taking 
linear combinations of the three physically relevant solutions of
\begin{equation}
\frac{\mathrm{d}^3}{\mathrm{d}x^3} (x(x-1))^{\frac{4}{3}}\frac{\mathrm{d}}{\mathrm{d}x} (x(x-1))^{\frac{2}{3}}\frac{\mathrm{d}}{\mathrm{d}x} F(x)\,,
\end{equation}
where $x$ is the crossing ratio and $F$ the conformally mapped crossing 
probability. The equation factorizes into \cite{Kleban:2002pf}
\begin{equation}
\left(\frac{\mathrm{d}^2}{\mathrm{d}x^2}(x(x-1)) + \frac{1}{2x-1}\frac{\mathrm{d}}{\mathrm{d}x}(2x-1)^2 \right)\frac{\mathrm{d}}{\mathrm{d}x} (x(x-1))^{\frac{1}{3}}\frac{\mathrm{d}}{\mathrm{d}x} (x(x-1))^{\frac{2}{3}}\frac{\mathrm{d}}{\mathrm{d}x} F(x)\,,
\end{equation}
where the rightmost part already provides us with the three expected 
solutions for the crossing probabilities in percolation.

This third order differential equation has neither direct interpretation 
as a third level null vector in a $c=0$ theory (more precisely: there is 
no such vector in this theory), nor does it arise from $h=0$ boundary 
operators. In contrary, we will show, that we obtain it from the null 
vector of an $h_{(1,3)}=-\frac{2}{3}$ field acting on a correlator containing 
$h_1 = h_2 = h_{(1,3)}=-\frac{2}{3}$ and $h_3 = h_{(1,5)}= -1$ in an $c_{(6,1)} = -24$
LCFT, which, for a level three 
null vector condition, is a unique solution.

The level 3 null vector can be written as \cite{Moghimi-Araghi:2000qn}
$$
  |\chi^{(3)}_h\ket = \left(L_{-1}^3 - 2(h+1)L_{-1}L_{-2} +
  (h+1)(h+2)L_{-3}\right)|h\ket\,.
$$
We will be a little bit more elaborate on this subject, since there are many 
errors in the equations found in the canonical literature (i.e. see 
\cite{DiFrancesco:1997nk} on pages 288).

Transforming this expression into a differential operator made out of the
${\cal L}_{-n}$ defined by
\begin{equation}\label{eq:Ln}
  {\cal L}_{-n}(z) = \sum_i\left(
    \frac{(n-1)h_i}{(z_i-z)^n} - \frac{1}{(z_i-z)^{n-1}}\partial_{z_i}\right)\,,
\end{equation}
acting on a 4-point function
\begin{equation}
  F(z,z_1,z_2,z_3) \equiv \langle\phi_h(z)\phi_{h_1}(z_1)\phi_{h_2}(z_2)
  \phi_{h_3}(z_3)\rangle\,,
\end{equation}
yields a quite lengthy expression. Replacing again
all derivatives $\partial_{z_i}$  by expressions only containing the
derivative $\partial_z$ with respect to $z$ and finally putting
$\{z_1,z_2,z_3\}\mapsto\{0,1,\infty\}$, results in the following ordinary
third order differential equation for $F(z)\equiv F(z,0,1,\infty)$:
\begin{eqnarray} 
  0 &=& \frac{{\rm d}^3}{{\rm d}z^3}F(z) 
		+ 2(h+1)\frac{2z-1}{z(z-1)}\frac{{\rm d}^2}{{\rm d}z^2}F(z)\nonumber\\
    &+& (h+1)\left( \frac{h-2 h_1}{z^2 } + \frac{h-2 h_2}{(z-1)^2 } 
		-2 \frac{h_3- h - h_1 - h_2}{z(z-1)}  + \frac{h}{z(z-1)} \right)
		\frac{{\rm d}}{{\rm d}z}F(z)\nonumber\\
    \label{eq:level3}
    &+& h(h+1) \left( -\frac{2h_1}{z^3 } - \frac{2h_2}{(z-1)^3 } 
		+ \frac{(2z-1)( h + h_1 + h_2 - h_3)}{z^2(z-1)^2} \right)F(z)\,.\phantom{mm}
\end{eqnarray}
Comparing this result to a simplified version of the differential equation
given by Watts \cite{Watts:1996yh}
\begin{equation}\label{eq:watts}
  \left(\frac{ {\rm d}^3}{ {\rm d}z^3} +
        \frac{5(2z-1)}{z(z-1)}\frac{ {\rm d}^2}{ {\rm d}z^2} +
        \frac{4}{3z(z-1)}\frac{ {\rm d}}{ {\rm d}z}\right)F(z) = 0\,.
\end{equation}
we know that this equation should be reproduced by (\ref{eq:level3}) for an
appropriate choice of $h,h_1,h_2,h_3$. However, (\ref{eq:watts}) does
not possess a term proportional to $F$ itself (not to one of its
derivatives). Clearly, in this form, this could only be the case for
$h=0$ or $h=-1$. One can easily see, that there are no triples
$\{h_1,h_2,h_3\}$ for these values of $h$ such that (\ref{eq:level3}) 
becomes equivalent to (\ref{eq:watts}). But there is a simple and
natural way out, since we know something about the generic form of a 4-point
function of four primary fields. For example, any function
$F(z,0,1,\infty)$, which is invariant under global conformal transformations,
must be of the form $F(z) = z^{\mu_{01}}(z-1)^{\mu_{02}}f(z)$. 
Using such an ansatz in (\ref{eq:level3}) and pulling the differential
operators through the pre-factor yields a modified differential equation for
$f(z)$. Nicely, $f(z)$ satisfies exactly (\ref{eq:watts}), if we put
$h=h_1=h_2=-2/3$ and $h_3=-1$. This implies $c=-24$, since then
the representation with highest weight $h=-2/3$ indeed possesses a null vector 
at level 3. Furthermore, the exponents $\mu_{01}=\mu_{02}=1/3$ are
exactly what one expects from the generic solution
$\mu_{ij}=\frac{1}{3}\sum_k h_k - h_i - h_j$ of 
$\sum_{j\neq i}\mu_{ij} = -2h_i$, 
i.e.~$(-2/3-2/3-2/3-1)/3 + 2/3 + 2/3 = -1 + 4/3 = 1/3$. To summarize, the
conformal blocks of the 4-point function 
\begin{equation}
\langle\Phi_{h=-2/3}(z)
\Phi_{h_1=-2/3}(0)\Phi_{h_2=-2/3}(1)\Phi_{h_3=-1}(\infty)\rangle 
= z^{\mu_{01}}(1-z)^{\mu_{02}} f(z) 
\end{equation}
of the $c=-24$ theory are in one-to-one correspondence with the
solutions of Watts' differential equation.

As a concluding remark we note that $h=-2/3$ corresponds to a reducible but
indecomposable representation of the $c_{(6,1)}=-24$ theory. Hence, it
is natural and inevitable, that correlation functions involving more than
one field of this type will contain conformal blocks with logarithmic
divergences. Indeed, the Watts fifth-order differential equation has 
three solutions plus two with logarithmic divergences. Thus a solution
from the augmented minimal models of the type $c_{(3p,3q)}$ is not
surprising and it seems to be an interesting application for LCFTs
\cite{Gurarie:1993xq,Gaberdiel:2001tr,Flohr:2001zs}. The logarithmic
behavior of such disorder models has already been conjectured before
\cite{davis-2000-570} thus the solution fits well into the general
expectations. Additionally we should mention that the
third-order equation has three regular solutions, which is in agreement 
with the fusion rules of this logarithmic CFT, where the irreducible
sub-representation with highest weight $h=-2/3$ satisfies
$[-2/3]*[-2/3]=[0]+[-2/3]+[1]$.
Further details will be worked out in a future publication.

Additionally, the field content of the $c=-24$ theory has a very interesting
property. Taking a look at 
the relevant entries of the Kac-Table (the first row is sufficient here)
$$
\begin{array}{|c|c|c|c|c|c|c|c|c|c|c|c|c|c|c|c|c|}
\hline
\rule[-.6em]{0cm}{1.6em}0 & 
-\frac{3}{8}  & 
-\frac{ 2}{3} & 
-\frac{ 7}{8} & 
-1 & 
-\frac{ 25}{24}&
-1 & 
-\frac{ 7}{8} & 
-\frac{ 2}{3} & 
-\frac{3}{8}  & 
0 &
\frac{ 11}{24}&
1 &
\frac{13}{8}  & 
\frac{ 7}{3} & 
\frac{25}{8}  & 
4 \\
\hline
\end{array} 
$$
we encounter that the critical exponents that are assumed to come up in 
percolation appear shifted by $1$, i.e. $h_{(1,2)}= - \frac{3}{8}$ and 
$h_{(1,4)}= - \frac{7}{8}$. More precisely, the effective conformal
weights $h_{{\rm eff}}=h-c/24$ agree, 
i.e.~$h_{{\rm eff}}^{c=-24}=h_{{\rm eff}}^{c=0}$.
Thus descendants of those fields could describe 
the physical properties of percolation. Additionally, the pre-logarithmic
field with conformal weight $h_{(0,0)}=h_{(1,6)}=-\frac{25}{24}$ appears,
in whose fusion product with itself the indecomposable representations
arise
\cite{Flohr:1995ea,Flohr:1996vc,Kogan:1997fd}

Further support for our conjecture that the rational logarithmic conformal
field theory with central charge $c=c_{(6,1)}=-24$ might describe percolation
is given by the following remarkable observation. The partition function of
this theory is equivalent to the partition function eq.~(\ref{eq:Zfull}) of 
the extended $c=0$ theory discussed above. More precisely, we have 
\cite{ Flohr:1995ea,Flohr:1996vc} that
\begin{equation}
  Z_{c_{(6,1)}=-24}[\alpha] = Z_{{\rm full}}[\alpha,\beta=0]\,.
\end{equation}
Therefore, the non-logarithmic parts of the two partition functions, 
which actually count the states, are identical.

On the other hand, this is not entirely surprising. Many arguments, which 
favour a conformal field theory with vanishing central charge for the
description of two-dimensional percolation, rely on the modular properties
of the partition function. These properties cannot fix the central
charge uniquely, but only modulo $24$. Surely enough, $c_{(6,1)}=-24\equiv 0
=c_{(3,2)}$ mod $24$, and the effective central charges are equal to one for 
both theories.

If we still want to describe percolation as a $c=0$ theory and still do not want
to reject the interpretation of Watts' differential equation as a level three null vector, we
may construct a tensorized CFT consisting of the $c=-24$ and a $c=24$ part.
Therefore, any correlation function or field 
factorizes into two parts, one for each of the two CFTs, i.e.
 $\Phi_H(z) = \Phi_{h,c=-24}(z)\otimes \Phi_{H-h,c=+24}(z)$.
However, since the 4-point function 
$$
F_{c=-24}(z)=\bra\Phi_{-2/3}(z)\Phi_{-2/3}(0)
\Phi_{-2/2}(1)\Phi_{-1}(\infty)\ket_{c=-24}
$$ 
already yields as solutions the 
desired crossing probabilities, the second factor, 
$$
G_{c=+24}(z)=
\bra\Phi_{h}(z)\Phi_{h_1}(0)\Phi_{h_2}(1)\Phi_{h_3}(\infty)\ket_{c=+24}
$$
should be trivial. To make the picture perfect, we could try to
achieve
$$
  f(z) = F_{c=-24}(z)G_{c=+24}(z)\ \ \ \ \Longrightarrow\ \ \ \
  G_{c=+24}(z) = z^{-1/3}(z-1)^{-1/3}\,.
$$
The easiest way to get that result is to assume that $G(z)$ is, essentially,
a 3-point function $\bra\Phi_{1/3}(z)\Phi_{1/3}(0)\Phi_{1/3}(1)\mathbb{I}(\infty)
\ket_{c=+24}$. It remains to clarify whether such a correlator exists and
is non-vanishing in a $c=+24$ theory.

But what about the results already derived and proven consistent with numerical
simulations for $\Pi_h$ if percolation was described by a $c_{(1,6)}=-24$ theory?

As already mentioned above, the horizontal crossing probability is determined 
by a second order differential equation interpreted as a level two null vector 
condition arising from $\phi_{(1,2)}$ which has the weight 
$h = h_{(1,2)} = -\frac{3}{8}$ in this case of $c=-24$. 
Therefore we have to solve
\begin{equation}\label{eq:level2}
\left( \frac{3}{2(2h+1)} \frac{\mathrm{d}^2}{\mathrm{d}z^2} + \frac{2z-1}{z(z-1)} \frac{\mathrm{d}}{\mathrm{d}z} - \frac{h_1}{z^2} - \frac{h_2}{(z-1)^2} + \frac{h + h_1 + h_2 - h_3}{z(z-1)} \right) F(z) = 0.
\end{equation}
with the central charge related to $h$ via $c=2h(5-8h)/(1+2h)$.

>From the numerical simulation of Langlands et~al.~\cite{Langlands:1994} we know, 
that Cardy's formula \cite{Cardy:1991cm} for $\Pi_h$ derived from a level two null 
vector in a $c=0$ minimal model should be the outcome. Thus we know that $F(z)$ 
should be of the form $\left.\right._2 F_1 (1/3,2/3,4/3,z)$. A simple calculation 
yields $h_1=h_2=h_3=h_{(1,4)}=- \frac{7}{8}$ and 
$F_1(z) = (z(z-1))^{\frac{1}{4}} \cdot z^{\frac{1}{3}} \left.\right._2 F_1 (1/3,2/3,4/3,z)$ 
as well as $F_2 (z)=(z(z-1))^{\frac{1}{4}} $ as the second solution. Hence 
in comparison to Cardy, the crossing probability for percolation is given by 
their quotient $F_1/F_2$. Now our solution for $\Pi_h$ has exactly the same 
properties as described in \cite{Langlands:1994} and thus is zero for 
$z \rightarrow 1$ and one for $z \rightarrow 0$, as desired.\footnote{This means that if we consider a rectangle whose corners are mapped clockwise in decreasing order to the $z_i$ with $r:= (z_3-z_0)/(z_1-z_0)$, $r \rightarrow 0$ and $r \rightarrow \infty$, respectively. Note that $0 < z < 1$ and therefore the correct mapping on the upper complex plane is taking $z_0 \rightarrow z$, $z_1 \rightarrow 0$, $z_2 \rightarrow \infty$ and $z_3 \rightarrow 1$.} The normalization is obtained by considering the identity
\begin{equation}
\frac{3\Gamma \left(\frac{2}{3}\right)}{\Gamma^2 \left(\frac{1}{3}\right)} \left.\right._2 F_1 (1/3,2/3,4/3,z) = 1 - \frac{3\Gamma \left(\frac{2}{3}\right)}{\Gamma^2 \left(\frac{1}{3}\right)}(1-z)^{\frac{1}{3}} \left.\right._2 F_1 (1/3,2/3,4/3,1-z).
\end{equation}
Hence the correct normalization constant must be $\frac{3\Gamma \left(\frac{2}{3}\right)}{\Gamma^2 \left(\frac{1}{3}\right)}$.

This result is remarkable, since it contains the two fields for critical exponents in percolation, i.e.~$h_{(1,2)}= - \frac{3}{8}$ and $h_{(1,4)}= - \frac{7}{8}$ in the $c=-24$ theory.
Another important thing to be considered are the results of SLE for percolation,
 showing the equivalence of Cardy's formula and the results for $\kappa = 6$. 
 At first we have to state that the frequently cited proof of Smirnow \cite{Smirnov:1994}
 (or Dubedat \cite{Dubedat:2004} as well) 
 only holds for site percolation on a triangular lattice, and according to Smirnow and 
 Werner \cite{Smirnov:2001}, the method used in \cite{Smirnov:1994} can not be applied 
 directly to bond percolation on the square lattice as discussed in this paper. 
 The problem with a proof of bond percolation on the square lattice seems to lie
 within the properties of the hypergeometric functions which appear to be the solutions
 of the null vector differential equations. As noted by L.~Carleson (we found
this mentioned in \cite{Cardy:2001vr})
the horziontal
 crossing probability is proportional to
 $$
 \int_0^\eta \left(t (1-t) \right)^{-2/3} \mathrm{d} t
 $$
 which is exactly the Schwarz-Christoffel mapping from the upper half plane to a
 equilateral triangle. Thus, for this special lattice, $\Pi_h$ becomes very simple
 which has rigorously been proven by Smirnow as stated above. This problem has been
 referred to by him as ``It seems that $2\pi/3$ rotational symmetry enters in our
 paper not because of the specific lattice we consider, but rather manifests some
 symmetry laws characteristic to (continuum) percolation.'' For the same reason,
 Dubedat's proof of Watts' formula \cite{Dubedat:2004} is only true for the triangular 
 case, too. The connection between SLE and triangular symmetry has also been described
 by him (see \cite{dubedat-2003-8}).
  
Additionally, we should keep in mind that at one point in the derivation of 
the differential equation for the SLE$_\kappa$-process, namely where the 
identification of the evolution operator $\mathcal{A}$ with a level two null 
vector of a CFT is done \cite{Bauer:2002tf}, the assumption, 
that $h_{(1,2)}=0$ is made. This has consequences on the relation between the coefficients of 
 the differential equation ($\kappa$, $c$  and $h_{(1,2)}$) and the evolution operator, 
\begin{equation}
\mathcal{A}= -2L_{-2} + \frac{\kappa}{2}L_{-1}^2 \quad vs. \quad L_{-2} - \frac{3}{2(2h_{(1,2)}+1)}L_{-1}^2.
\end{equation}
Hence, we know that
\begin{equation}
\frac{\kappa}{4} = \frac{3}{2(2h_{(1,2)}+1)}.
\end{equation}
Obviously, this leaves us with $\kappa = 6$ if we restrict ourselves to $h=0$ 
in our ansatz for percolation (or equivalently $c_{(p,q)}=c_{(3,2)}=0$ which 
means $\frac{3}{2(2h_{(1,2)}+1)} = \frac{q}{p} = \frac{3}{2}$). 
But since there are no compulsory conditions to justify this ansatz as 
explained before, we may question why we should not try $h=-\frac{3}{8}$ and 
thus $\kappa = 24$ or $h=4$ and $\kappa = \frac{2}{3}$. We are aware of the
fact that a solution $\kappa = 24$ is problematic since for this value of
$\kappa$ the curve is space filling. Thus this can be a hint that two dimensional
critical bond percolation may have to be formulated in a more complicated setup
if it is described by a $c=-24$ LCFT. 

There is, however, one possibility to try to elucidate this question further.
In \cite{math/9911084}, a generalization of the SLE process
related to percolation is proposed, which yields generalized probabilities,
depending on a parameter $b$ and given by the formula
\begin{equation}\label{eq:pib}
  \Pi(b;z) = z^{b+\frac{1}{6}}{}_2F_1(\frac{1}{6}+b,\frac{1}{2}+b;1+2b;z)\,.
\end{equation}
Obviously, $b=1/6$ reproduces the case relevant for percolation, and thus
this is referred to as a generalization of Cardy's formula.
It is clear that (\ref{eq:pib}) cannot be given in terms of 4-point
functions for all values of $b$ for one and the same CFT with fixed central
charge, But we can try to check, whether (\ref{eq:pib}) can be reproduced
by 4-point functions of CFTs whose central charges $c$ depend on the choice $b$.
We restrict ourselves to the case of positive rational $b\in\mathbb{Q}$, 
$b=p/q$. We then further require that all four fields in the correlator
shall be degenerate primary fields, i.e.~have conformal weights $h_{(r,s)}(c)$
from the Kac-table.

Thus, we have to match the general solution of the second-order
differential equation (\ref{eq:level2}) for a level two null field with 
the desired expression (\ref{eq:pib}). This leads to the result
\begin{equation}
  F(z) = [z(1-z)]^{-\frac{2}{3}h}\Pi(b;z)\,,\ \ \ \ {\textstyle
  h_1 = \frac{36b^2 - (4h-1)^2}{24(2h+1)}\,,\ \ \ \
  h_2=h_3 = -\frac{h(2h-1)}{3(2h+1)}}\,,
\end{equation}
Now, $h=h_{(1,2)}$ is a member
of the Kac-table by construction, but we have to check, whether $h_1$ and
$h_2$ can also be chosen from the same Kac-table, since $c$ is already
fixed by the choice of $h$ via $c=2h(5-8h)/(1+2h)$. It will be convenient
to introduce the parametrization $c=c(t)=13-6(t+1/t)$. Let us assume that
$b=p/q>0$, and that $h_2=h_{(r,s)}$, $h_3=h_{(r',s')}$. Plugging $h=h_{(1,2)}$ into
the solutions for $h_1$ and $h_2$, and then solving for $s$ and $s'$, 
respectively, leads to the diophantine equations
\begin{equation}
  s=t\left(r \pm 2\frac{p}{q}\right)\,,\ \ \ \ s'=t\left(r'\pm\frac{1}{3}\right)
  \,,
\end{equation}
where the parameter $t$ is the one used to parametrize $c=c(t)$.
There are various solutions to this, but clearly $t={\rm lcm}(3,pq)$ and
thus $c=c_{(t,1)}$ will it always make possible to find positive $r,r',s,s'$ 
such that all conformal weights are from the Kac-table.

Finally, we observe that $F(z)$ is only proportional to the desired
quantity $\Pi(b;z)$. Again, we would like to have that the quotient of 
the two conformal blocks, or correlations functions,
gives the probability, $\Pi(b;z)=F_1(z)/F_2(z)$.
To this end, we would need that $F_1=F$ and that $F_2$ is of the simple form
$F_2(z)=[z(1-z)]^{-\frac{2}{3}h}$. This is possible, if the charge balance,
in a free field realisation of the CFT, adds up to the background charge,
such that no screening integrations, which lead to a non-trivial $F_2(z)$,
are necessary. This yields us a further condition, since the charges
are 
\begin{equation}
  \alpha_{r,s}=\frac{1}{2}(r-1)\sqrt{t} + \frac{1}{2}(1-s)\frac{1}{\sqrt{t}}
  \,,\ \ \ \
  \alpha_0 = \frac{1}{2}\left(\sqrt{t}-\frac{1}{\sqrt{t}}\right)\,.
\end{equation}
We must have $\alpha_{1,2}+\alpha_{r,s}+2\alpha_{r',s'}=2\alpha_0$.
There is no good solution to this, but one easily can check that
$\alpha_{1,2}+3\alpha_{r',s'}=2\alpha_0$ is automatically fulfilled.
We therefore arrive at the result that for all $b=p/q>0$, a logarithmic
CFT with central charge $c=c_{(t,1)}$, $t={\rm lcm}(3,pq)$, reproduces the
generalized version of Cardy's formula as follows: Since $t$ is always
divisible by three, we put $t=3t'$, $t'\in\mathbb{N}$, and have
\begin{equation}
  \Pi(b;z) = \frac{\langle
  \phi_{(1,2)}(z)\phi_{(1,3t'(1\pm 2b))}(0),\phi_{(1,2t')}(1)\phi_{(1,2t')}(\infty)
  \rangle}{\langle
  \phi_{(1,2)}(z)\phi_{(1,2t')}(0),\phi_{(1,2t')}(1)\phi_{(1,2t')}(\infty)
  \rangle}\,.
\end{equation}
Note that $3t'(1\pm 2b)$ is always an integer. For $b<1$, we can choose
the minus sign, otherwise, we should choose the plus sign. Both cases
are within the augmented Kac-table for the rational logarithmic
models with central charge $c_{(3t',1)}$.

Interestingly, the known solution for $b=1/6$ in terms of a CFT with 
$c_{(3,2)}=0$
cannot be extended in a unified fashion to a series of CFTs for all rational 
$b$. Although this is no rigorous proof, this result might indicate that
our proposal is more natural.
\section*{Comments on the relation of \boldmath{$c\!=\!0$} and \boldmath{$c\!=\!-24$}}
After having demonstrated how important quantities which can be derived within
a $c=0$ CFT can equally well be deduced within a $c=-24$ rational CFT ansatz, 
we may ask the question
how these two theories are connected besides their effective central charges
being the same, as stated above. Therefore let us take a look at the extended 
Kac tables for both models. 

$c_{(3\cdot 3,3 \cdot 2)}:$
\small
$$ \label{eq:Kac0}
\begin{array}{|c|c|c|c|c|c|c|c|} 
\hline
\rule[-.6em]{0cm}{1.6em}0& 0 &\frac{1}{3}&1 &2 &\frac{10}{3}& 5 & 7\\
\hline
\rule[-.6em]{0cm}{1.6em} \frac{5}{8} & \frac{1}{8} & -\frac{1}{24} & \frac{1}{8} & \frac{5}{8} &\frac{35}{24} & \frac{21}{8} & \frac{33}{8}\\
\hline
\rule[-.6em]{0cm}{1.6em} 2 & 1 & \frac{1}{3} & 0 & 0 & \frac{1}{3} & 1 & 2\\
\hline
\rule[-.6em]{0cm}{1.6em} \frac{33}{8} & \frac{21}{8} & \frac{35}{24} & \frac{5}{8} & \frac{1}{8} & -\frac{1}{24} & \frac{1}{8} & \frac{5}{8} \\
\hline
\rule[-.6em]{0cm}{1.6em} 7 & 5 & \frac{10}{3} & 2 & 1 & \frac{1}{3} & 0 & 0 \\
\hline
\end{array}
$$
\normalsize
and $c_{(3 \cdot 6 ,3 \cdot 1)}:$
$$\label{eq:Kac-24}\!\!\!\!\!\!\!\!\!\!\!
\begin{array}{|c|c|c|c|c|c|c|c|c|c|c|c|c|c|c|c|c|}
\hline
\rule[-.6em]{0cm}{1.6em}0 & 
-\frac{3}{8}  & 
-\frac{ 2}{3} & 
-\frac{ 7}{8} & 
-1 & 
-\frac{ 25}{24}&
-1 & 
-\frac{ 7}{8} & 
-\frac{ 2}{3} & 
-\frac{3}{8}  & 
0 &
\frac{ 11}{24}&
1 &
\frac{13}{8}  & 
\frac{ 7}{3} & 
\frac{25}{8}  & 
4 \\
\hline
\rule[-.6em]{0cm}{1.6em}4 & 
\frac{25}{8}  & 
\frac{ 7}{3} & 
\frac{13}{8}  & 
1 &
\frac{ 11}{24}&
0 &
-\frac{3}{8}  & 
-\frac{ 2}{3} & 
-\frac{ 7}{8} & 
-1 & 
-\frac{ 25}{24}&
-1 & 
-\frac{ 7}{8} & 
-\frac{ 2}{3} & 
-\frac{3}{8}  &
0\\
\hline
\end{array}
$$
\normalsize
Obviously, not by multiplicity but by weight, all fields of the $c=0$ theory
are present in the $c=-24$ as well if we shift them by $-1$. As already
mentioned, the sets of effective conformal weights are thus equivalent. 
The similarities go further with remarkable consequences when
we consider differential equations due to null vectors. For instance,
let us take the level two case. For any choice of the other three fields 
$(X,Y,Z)$ in the four point function of $c=0$ Kac table fields,
\begin{equation}
\bra h_{(1,2)}^{0} X Y Z \ket \text{ or } \bra h_{(2,1)}^{0} X Y Z \ket\,, 
\end{equation}
we can find corresponding weights (X',Y',Z') in the Kac table of $c=-24$ 
such that there are corresponding four-point functions
\begin{equation}
\bra h_{(1,2)}^{-24} X' Y' Z' \ket \text{ or } \bra h_{(2,1)}^{-24} X' Y' Z' \ket\,,
\end{equation}
which yield the same solutions with respect to the action of the level two
null vector operator
\begin{equation}
\frac{3}{2(2h+1)}\mathcal{L}_{-1}^2 - \mathcal{L}_{-2}.
\end{equation}
Since the computation is very easy, we will only state the results:
Choosing for the fields $X,Y,Z$ any of the weights $H$, we have to
choose for the fields $X',Y',Z'$ the correpsonding weights $H'$ as given in the
two following tables.
\begin{equation}
h_{(1,2)}:
\begin{array}{|l||c|c|c|c|c|}
\hline
H & 1/8 & 0 & -1/24 & 5/8 & 1\\
\hline
H' & -3/8 & -7/8 & -25/24 & 13/8 & 25/8\\
\hline
\end{array}
\end{equation}
and analogously
\begin{equation}
h_{(2,1)}:
\begin{array}{|l||c|c|c|c|c|c|}
\hline
H & -1/24 & 1/8 & 5/8 & 33/8 & 21/8 & 35/24\\
\hline
H' & -25/24 & -1 & -7/8 & 0 & -3/8 & -2/3\\
\hline
\end{array}
\end{equation}
with $h_{(1,2)}$ being $h=0$ and $h=-3/8$ and $h_{(2,1)}$ being $h=5/8$ and $h=4$
for $c=0$ and $c=-24$, respectively. Thus it is not surprising that Cardy's formula
\cite{Cardy:1991cm} has also a meaning in $c=-24$.

Additionally, the structure of the Jordan cells of rank two
within the two LCFTs is very similar, for any non integer weight
we can find triplets corresponding to an irreducible representation
which is contained in an indecomposable of the same weight which
is isomorphic (with respect to the counting of states) to a
hidden indecomposable representation whose subrepresentation is
present in the Kac table and is based on a highest weight differing
by an integer from the two other triplet members. Details on this 
structure can be found explained within the famous $c=-2$ LCFT
example \cite{Gaberdiel:1996np,Gaberdiel:1998ps,Kausch:2000fu}. It is
present in the $c_{(t,1)}$ series of LCFTs \cite{Flohr:1995ea} and is
conjectured to exist in all augmented minimal models \cite{Flohr:1996vc}.
In the present case, we find such triplets for $c_{(3,2})=0$ and 
$c_{(6,1)}=-24$ respectively as
\begin{eqnarray}
(5/8,5/8,21/8) &\leftrightarrow& (-3/8,-3/8,13/8)\\
(1/3,1/3,10/3) &\leftrightarrow& (-2/3,-2/3,7/3)\\
(1/8,1/8,33/8) &\leftrightarrow& (-7/8,-7/8,25/8)
\end{eqnarray}
Unfortunately, the structure of the integer weights (or, more precisely
the weights that have previously been inside the Kac table of the non
augmented minimal model) can not be revealed by this analogy since
they are assumed to reside in a Jordan cell of rank three \cite{Holger}
which is known not to appear in $c_{(p,1)}$ models. Research on the details
is currently
going on heading towards a clarification of the representation structure
of $c_{(9,6)}=0$ which will provide us with the necessary knowledge
to establish a well-founded link between the two LCFTs rather than
just educated guesswork.

\section{Conclusion and perspective}
In this paper, we have shown that if we want to describe two dimensional bond percolation within a conformal field theory, using a level three null vector condition to get a differential equation for horizontal-vertical crossing probability $\Pi_{hv}$ that fits the numerical data, we have to take $c=-24$. This solution is unique. 

Additionally, there are no strict arguments contradicting our result, even not from the derivation of the horizontal crossing probability $\Pi_h$ whose form has already been proven in the literature, since it can be explained in our $c=-24$ CFT proposal as well. Hence the question remains if we should consider percolation being rather a $c=-24$ than the commonly assumed $c=0$ theory. Although we have presented several arguments
indicating that our proposal is more natural, and that some arguments in
favour of the $c=0$ theory are problematic (particularly the serious
issue of having a partition function $Z=1$ and simultaneously a $\phi_{(1,3)}$
field in the spectrum of the $c=0$ theory), we do not have a strict proof
for our solution.

But there are still open questions that arise when considering SLE. Of which we will
ask the perhaps most important one: 
Is there a (generalized) SLE corresponding to bond percolation on the square lattice? 
If yes, what are its properties? Is the proof explicit in both directions? 
Does it endorse or destroy the ansatz of $c=-24$?

Besides the discussion whether one or the other ansatz is correct, another 
important issue is to investigate in more detail the close relationship between
conformal field theories whose central charges differ by multiples of $24$,
especially why $c=-24$ and $c=0$ have so many similar properties concerning 
percolation. This question will be pursued in a forthcoming publication
\cite{Holger}.

\nocite{Ziff:1995a}
\nocite{Ziff:1995b}
\small
\bibliography{letterbib2}
\end{document}